\documentclass[pra,twocolumn,amsmath,amssymb,floatfix,reprint,footinbib,superscriptaddress,longbibliography,showkeys]{revtex4-2}
\usepackage[english]{babel}
\usepackage{picinpar,graphicx}
\usepackage{braket}
\usepackage{amsmath}
\usepackage{graphicx}
\usepackage{epstopdf}
\usepackage{dcolumn}
\usepackage{graphicx}
\usepackage{mathrsfs}
\usepackage{mdwlist}
\usepackage{subfigure}
\usepackage{booktabs}
\usepackage{amsmath}
\usepackage{dsfont}
\usepackage{amstext}
\usepackage{amssymb}
\usepackage{amsbsy}
\usepackage{bbm}
\usepackage{amsthm}
\usepackage{graphicx}
\usepackage{textcomp}
\usepackage{color}
\usepackage[colorlinks,citecolor=blue]{hyperref}
\usepackage{lipsum}
\definecolor{Dgreen}{RGB}{0, 100, 0}
\usepackage{url}
\usepackage[colorlinks]{hyperref}
\hypersetup{%
	plainpages=true,
	breaklinks=true,  
	hypertexnames=false,  
	pageanchor=true,
	colorlinks=true,
	linkcolor={blue},
	citecolor={red},
	urlcolor={blue},
	anchorcolor={black}
}

\begin{document}
	
	\title{Controllable Non-Hermitianity in Continuous-Variable Qubits}
	\author{Ke-Xiong Yan}
	\affiliation{Fujian Key Laboratory of Quantum Information and Quantum Optics, Fuzhou University, Fuzhou 350116, China}
	\affiliation{Department of Physics, Fuzhou University, Fuzhou, 350116, China}
	
	\author{Zhi-Cheng Shi}
	\affiliation{Fujian Key Laboratory of Quantum Information and Quantum Optics, Fuzhou University, Fuzhou 350116, China}
	\affiliation{Department of Physics, Fuzhou University, Fuzhou, 350116, China}
	
	\author{Ye-Hong Chen}\thanks{yehong.chen@fzu.edu.cn}
	\affiliation{Fujian Key Laboratory of Quantum Information and Quantum Optics, Fuzhou University, Fuzhou 350116, China}
	\affiliation{Department of Physics, Fuzhou University, Fuzhou, 350116, China}
	
	\affiliation{Quantum Information Physics Theory Research Team, Center for Quantum Computing, RIKEN, Wako-shi, Saitama 351-0198, Japan}
	
	\author{Yan Xia}\thanks{xia-208@163.com}
	\affiliation{Fujian Key Laboratory of Quantum Information and Quantum Optics, Fuzhou University, Fuzhou 350116, China}
	\affiliation{Department of Physics, Fuzhou University, Fuzhou, 350116, China}

	\date{\today}
	\begin{abstract}
	 Pure dephasing is the dominant leak mechanism in photonic cat qubits because its phase errors disrupt the parity protection, rendering the qubit vulnerable to energy relaxation. In this manuscript, we reveal that this dephasing mechanism conceals an interesting physical phenomenon: it induces \textit{asymmetric leakage} from the cat-state subspace, where even- and odd-parity cat states decay at different rates. This leak asymmetry enables the dynamics of the system to be described by a non-Hermitian Hamiltonian, thereby transforming the cat qubit into a platform with controllable gain and loss for probing non-Hermitian physics. Within this platform, we demonstrate the possibility to control the parity-time symmetry phase transition in a single cat qubit by adjusting its amplitude. Moreover, we couple two cat qubits to realize an entanglement phase transition induced by the exceptional point. Our work constructs a controllable non-Hermitian system simulator, overturning the conventional paradigm that treats dephasing as harmful noise.
	\end{abstract}
	
	\maketitle

	\textit{Introduction.}---Continuous-variable systems, encoding logical qubits in the continuous variables of bosonic modes~\cite{PhysRevA.59.2631,PhysRevA.64.012310,PhysRevA.73.012325,PhysRevLett.111.120501,PhysRevX.6.031006,PhysRevA.97.032346,CampagneIbarcq2020,Ma2021,Sivak2023} (e.g., the amplitudes of light fields), have attracted growing interest in recent years~\cite{Hu2019,Ma2020,Cai2021}. Owing to the infinite-dimensional Hilbert space of bosonic modes~\cite{RevModPhys.77.513,Andersen2010,RevModPhys.84.621,Andersen2015,Pfister2019,Slussarenko2019,Fukui2022}, a logical qubit can be encoded by extending only the number of excitation instead of the number of qubits, making the continuous-variable encodings inherently hardware-efficient~\cite{Chiaverini2004,PhysRevLett.102.070502,Schindler2011,PhysRevA.86.032324,PhysRevA.97.022335,PhysRevApplied.13.014055}. Among various continuous-variable-encoded qubits, the cat qubit is particularly intriguing~\cite{Mirrahimi2014,Leghtas2015,Puri2017,PhysRevLett2021,PRXQuantum.3.010329,PhysRevResearch.4.013082}. Its logical states are superpositions of two coherent states with opposite phases, i.e., $\ket{\mathcal{C_{\pm}}}=\mathcal{N_{\pm}}(\ket{\alpha}\pm \ket{-\alpha})$ (where $\mathcal{N_{\pm}}$ are their normalization coefficients and $\alpha$ is the amplitude of the coherent state). A key feature of this encoding is its inherent noise bias~\cite{Guillaud2023,Rglade2024}: as the amplitude of the coherent states (the ``cat size'') increases, the bit-flip error rate is exponentially suppressed, whereas the phase-flip error rate increases only linearly.
	
    The realization of a cat qubit requires confining the system dynamics to the cat manifold, which relies on a delicate interplay of engineered parametric processes~\cite{PRXQuantum.4.020337}. In recent years, two primary approaches have been proposed to achieve such confinement: (i) \textit{Dissipative} cat qubit utilizes an engineered dissipation scheme that combines a two-photon drive and two-photon loss to generate and autonomously stabilize the code manifold~\cite{Mirrahimi2014,Leghtas2015,PhysRevX.8.021005,Lescanne2020}. This architecture is intrinsically robust against leakage processes and ensures exponential suppression of the bit-flip error rate with the cat size~\cite{Mirrahimi2014,Leghtas2015,PhysRevX.6.041031,PhysRevA.97.032346}; (ii) \textit{Kerr-cat} qubit, by contrast, utilizes Kerr nonlinearity to restrict the system to the doubly degenerate ground space of a Kerr parametric oscillator (KPO)~\cite{Puri2017,PhysRevX.9.041009,Puri2020,Grimm2020}. This configuration can be effectively modeled as a double-well potential, where the tunneling rate between the wells is exponentially suppressed with the cat size $|\alpha|^2$~\cite{PhysRevX.9.041053}. Both architectures have been successfully implemented in
    superconducting circuits~\cite{Puri2017,Leghtas2015,PhysRevX.8.021005,Grimm2020,Ofek2016,PRXQuantum.2.030345,PhysRevX.14.041049,PhysRevX.15.011070}, establishing cat qubits as a leading platform for bosonic quantum computing.
	
	In addition to the role in quantum error correction, cat qubits have demonstrated remarkable capabilities in quantum simulation recently due to their noise bias characteristics~\cite{PhysRevLett.133.033603}. In this manuscript, we further demonstrate that the phase noise induced by photon number fluctuations enables the cat qubit to serve as a well-controlled platform for simulating non-Hermitian physics. Specifically, when the system dynamics are restricted to the subspace spanned by even- and odd-parity cat states, the photon number fluctuation operator causes the two parity cat states to leak into states outside the subspace at different rates. Such an \textit{asymmetric leakage} results in continuous loss of population from the cat-state subspace $\{\ket{\mathcal{C_{\pm}}}\}$, thereby causing the system dynamics to be governed by a non-Hermitian Hamiltonian. The imaginary components of this non-Hermitian Hamiltonian directly reflect the rates of such \textit{asymmetric leakage}.

	We note that non-Hermitian dynamics has enabled remarkable capabilities in enhanced parameter sensing~\cite{PhysRevLett.112.203901,PhysRevLett.117.110802,PhysRevA.93.033809,Hodaei2017,Chen2017,Lau2018,PhysRevLett.123.180501,PhysRevLett.125.240506,McDonald2020,PhysRevA.103.042418,PhysRevResearch.6.023216}, quasi-adiabatic evolution chiral mode switching~\cite{Uzdin2011,Berry2011,PhysRevA.88.033842,PhysRevA.89.033403,PhysRevA.92.052124,Doppler2016,Xu2016,PhysRevA.102.032216,PhysRevLett.133.113802}, directional invisibility~\cite{PhysRevLett.106.213901,PhysRevLett.131.036402}, and robust wireless power transfer~\cite{Assawaworrarit2017}. To further explore the unique phenomena and applications of non-Hermitian physics, there exists a pressing need to develop flexible and robust physical platforms for realizing such dynamics. Our work \textit{bridges the fields of fault-tolerant encoding and non-Hermitian physics, showcasing how engineered dissipation can transform noise into a quantum simulation resource}. To exemplify this, we simulate two non-Hermitian phenomenons: (i) By controlling the size $\alpha$ of the cat state, we observe a phase transition associated with parity-time ($\mathcal{P}\mathcal{T}$) symmetry breaking in a cat-state subspace $\{\ket{\mathcal{C_{\pm}}}\}$; (ii) Extending to two coupled cat qubits, we achieve an entanglement phase transition induced by an exceptional point, i.e., a phenomenon with no direct counterpart in Hermitian quantum systems.

	\textit{Pure-dephasing-induced asymmetric leakage.}---Since coherent states are eigenstates of the photon annihilation operator $\hat{a}$, it is trivial to see that $\hat{a}\ket{\mathcal{C}_{\pm}}=\alpha p^{\pm1}\ket{\mathcal{C}_{\mp}}$, where $p=p^{+1}=\mathcal{N}_{+}/\mathcal{N}_{-}$~\cite{Scully1997}. Therefore, in the cat-state subspace $\{\ket{\mathcal{C_{\pm}}}\}$, the annihilation operator induces transitions between even- and odd-parity cat states. However, the coherent states are not eigenstates of the photon creation operator, i.e., $\hat{a}^{\dagger}\ket{\alpha}=\alpha\ket{\alpha}+D(\alpha)\ket{1}$ and $\hat{a}^{\dagger}\ket{-\alpha}=-\alpha\ket{-\alpha}+D(-\alpha)\ket{1}$, where $\ket{1}$ is the Fock state and $D(\pm\alpha)=\exp[\pm\alpha(\hat{a}^{\dagger}-a)]$ are the displace operators. The above result implies that the actions of $\hat{a}^{\dagger}$ and $\hat{a}^{\dagger}\hat{a}$ on coherent states $\ket{\pm \alpha}$ or cat states $\ket{\mathcal{C_{\pm}}}$ can cause leakage out of the code space. Therefore, dephasing processes also lead to population leakage from the cat-state subspace $\{\ket{\mathcal{C_{\pm}}}\}$. More importantly, the pure-dephasing channel induced by the photon number operator acts asymmetrically on the two parity sectors within the cat-state subspace $\{ \ket{\mathcal{C}_{\pm}}\}$. This leads to a parity-dependent loss of population from the subspace. From the viewpoint of an effective dynamics for the subspace, this dissipative process can be captured by a non-Hermitian Hamiltonian, where the imaginary parts of its eigenvalues correspond to the distinct leakage rates of the logical basis states.
	
\begin{figure}
		\centering
		\includegraphics[scale=0.98]{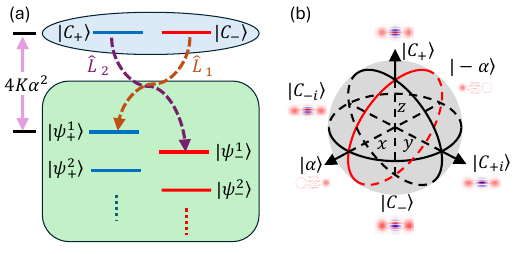}
		\caption{(a) Energy spectrum of the Kerr parametric oscillator. The eigenstates separate into even- and odd-parity manifolds, with the cat-state subspace $\{\ket{\mathcal{C}_{\pm}}\}$ forming the ground states. The excited states appear at a lower energy because the Kerr nonlinearity is negative. The energy gap to the first excited states scales as $\omega_{\mathrm{gap}}\simeq4K\alpha^2$, providing protection against unwanted transitions. The purple and brown dashed arrows in (a) indicate the leakage from the cat-state subspace $\{\ket{\mathcal{C_{\pm}}}\}$ due to pure dephasing. (b) Bloch sphere of the cat qubit and the corresponding Wigner function representations.}
		\label{fig1}
	\end{figure}

	In this manuscript, we employ a KPO to illustrate the non-Hermitian dynamics arising from such \textit{asymmetric leakage}. The Hamiltonian in a frame rotating at the resonator frequency is 
	\begin{equation}
		\hat{H}_{\rm KPO}=-K\hat{a}^{\dagger}\hat{a}^{\dagger}\hat{a}\hat{a}+P(\hat{a}^{\dagger2}+\hat{a}^{2}).
		\label{eq2}
	\end{equation} 
	In the above expression, $K$ is the amplitude of the Kerr nonlinearity and $P$ is the amplitude of the two-photon drive. The corresponding cat size is $\alpha=\sqrt{P/K}$~\cite{Grimm2020}. Figure~\ref{fig1}(a) shows the energy spectrum of the KPO. Its eigenstates are composed of two distinct parity classes. Within this structure, the even and odd cat states form a degenerate cat-state subspace $\{ \ket{\mathcal{C}_{\pm}}\}$, which can be employed as the computational basis states, as illustrated in Fig.~\ref{fig1}(b). The other eigenstates $|\psi_{\pm}^{k}\rangle$ (where $k$ indicates the $k$-th excitation manifold) with their respective normalization coefficients $\mathcal{N}_{\pm}^{k}$, constitute a excited-state subspace that is orthogonal to the cat-state subspace $\{\ket{\mathcal{C_{\pm}}}\}$. A key feature of the KPO is the large and tunable energy gap $\omega_{\mathrm{gap}}\simeq4K\alpha^2$, which separates the computational subspace $\{ \ket{\mathcal{C}_{\pm}}\}$ from the first excited states $|\psi_{\pm}^{1}\rangle$~\cite{PhysRevX.9.041009,PhysRevApplied.18.024076}.
	
	The evolution of the system under pure dephasing is governed by the Lindblad master equation:
	\begin{equation}
		\dot{\hat{\rho}}=-i[\hat{H}_{\mathrm{KPO}},\hat{\rho}]+\kappa^{\phi}\mathcal{D}[\hat{a}^{\dagger}\hat{a}]\hat{\rho},
		\label{eq3}
	\end{equation}
	where $\mathcal{D}[\mathcal{\hat{A}}]\hat{\rho}=\mathcal{\hat{A}}\rho\mathcal{\hat{A}}^{\dagger}-1/2(\mathcal{\hat{A}}^{\dagger}\mathcal{\hat{A}}\rho+\rho \mathcal{\hat{A}}^{\dagger}\mathcal{\hat{A}})$ and $\kappa^{\phi}$ is the pure dephasing rate of the cavity mode. The dephasing operator $\hat{a}^{\dagger}\hat{a}$ representing photon number fluctuations, induces decoherence within the cat-state subspace $\{\ket{\mathcal{C_{\pm}}}\}$ and population leakage out of it. When the second term on the right-hand side of Eq.~(\ref{eq3}) is projected onto the eigenstates space of the Hamiltonian $\hat{H}_{\mathrm{KPO}}$, we find that the dephasing superoperator connects the cat-state subspace $\{\ket{\mathcal{C_{\pm}}}\}$ and orthogonal partner, as illustrated in Fig.~\ref{fig1}(a). This connection is captured by two distinct leakage operators~(See the Supplementary Material~\cite{YKX251014} for details):  $\hat{L}_{1}=\sqrt{p\mathcal{N}_{+}/\mathcal{N}_{+}^{1}}|\psi_{+}^{1}\rangle\langle\mathcal{C}_{-}|$ and
	$\hat{L}_{2}=\sqrt{p^{-1}\mathcal{N}_{-}/\mathcal{N}_{-}^{1}}|\psi_{-}^{1}\rangle\langle\mathcal{C}_{+}|$. These operators define two \textit{asymmetric leakage} channels: one from the odd cat state $\ket{\mathcal{C}_-}$ to the even-parity excited state $\ket{\psi_+^{1}}$, and the other from the even cat state $\ket{\mathcal{C}_+}$ to the odd-parity excited state $\ket{\psi_{-}^{1}}$, each occurring at a different rate.
	
	 Therefore, when the system is initialized within the cat-state subspace $\{\ket{\mathcal{C_{\pm}}}\}$, the dissipative dynamics can be approximated by an effective non-Hermitian Hamiltonian. This is derived by incorporating the anti-Hermitian part of the Lindbladian into the Hamiltonian of the system, leading to~(See in the Supplementary Material~\cite{YKX251014}): 
	\begin{equation}
		\dot{\hat{\rho}}=-i\left[(\hat{H}_{\rm KPO}-\frac{i\kappa^{\phi}\alpha^{2}}{2}\sum_{n=1,2}\hat{L}_{n}^{\dagger}\hat{L}_{n}),\ \ \hat{\rho}\right].
		\label{eq9}
	\end{equation}
	Calculating the operator products $\hat{L}_{n}^{\dagger}\hat{L}_{n}$ yields a result: their action is confined to the cat-state subspace $\{ \ket{\mathcal{C_{\pm}}}\}$, giving rise to a purely imaginary diagonal Hamiltonian:
	\begin{equation}
		\left.\hat{H}_{\mathrm{eff}}=\left(
		\begin{array}
			{cc}i\gamma & 0 \\
			0 & -i\gamma
		\end{array}\right.\right).
	\end{equation}
	Here, the coefficient 
	\begin{equation}
		\gamma=\frac{\kappa^{\phi}\alpha^{2}}{2}\left[\frac{p^{-1}\mathcal{N}_{-}}{\mathcal{N}_{-}^{1}}-\frac{p\mathcal{N}_{+}}{\mathcal{N}_{+}^{1}}\right]
	\end{equation}
	quantifies the \textit{asymmetry in leakage rates} between the even and odd cat states. This \textit{asymmetry} is the fundamental origin of the non-Hermitian dynamics, effectively acting as a parity-dependent gain or loss.
	
	To probe the resulting non-Hermitian dynamics, we introduce a weak single-photon drive, $H_{c}=\Omega(\hat{a}^\dagger +\hat{a})$, with $\Omega \ll \omega_{\mathrm{gap}}$ to avoid exciting states outside the computational subspace.
	The total effective Hamiltonian in the cat-state subspace ${\ket{\mathcal{C_{\pm}}}}$ becomes~(See in the Supplementary Material~\cite{YKX251014})
	\begin{equation}
		\hat{\mathcal{H}}=
		\begin{pmatrix}
			i\gamma & \epsilon \\
			\epsilon & -i\gamma
		\end{pmatrix},
		\label{eq14}
	\end{equation}
	where $\epsilon=2\alpha \Omega$. This is the $\mathcal{P}\mathcal{T}$-symmetric dimer, a paradigmatic model in non-Hermitian physics~\cite{ElGanainy2018}.
	
	\begin{figure}
		\centering
		\includegraphics[scale=0.98]{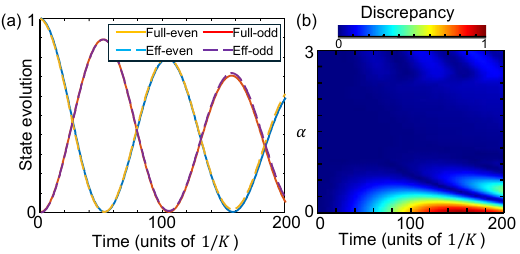}
		\caption{(a) Time evolution of the even cat state and the odd cat state governed by the effective  master equation in Eq.~(\ref{eq9}) (dashed curve) and the full master equation in Eq.~(\ref{eq3}) (solid curve). Here, we choose $\alpha=1.5$.  (b) Discrepancy in the population of the even cat state under the effective and full master equations, as a function of the cat state amplitude $\alpha$ and time. All simulations include a weak single-photon drive with amplitude $\Omega=0.01K$. Other parameters is $\kappa^{\phi}=0.001K$. }
		\label{fig2}
	\end{figure}
	
    To validate our effective model in Eq.~(\ref{eq9}), we compare it with numerical simulations of the full master equation in Eq.~(\ref{eq3}). Figure~\ref{fig2}(a) shows the populations of the even and odd cat states for $\alpha=1.5$. The agreement between the effective (dashed curves) and full (solid curves) dynamics confirms that the non-Hermitian Hamiltonian $\hat{\mathcal{H}}$ faithfully captures the essential dynamics. To quantify the accuracy over a broad parameter range, Fig.~\ref{fig2}(b) shows the discrepancy in the population of the even cat state. The error remains below $1\%$ for $\alpha \in [1, 2.5]$, demonstrating the robustness of our effective description and establishing the Kerr cat qubit under dephasing as a highly controllable simulator of non-Hermitian quantum mechanics.

\begin{figure}
		\centering
		\includegraphics[scale=0.98]{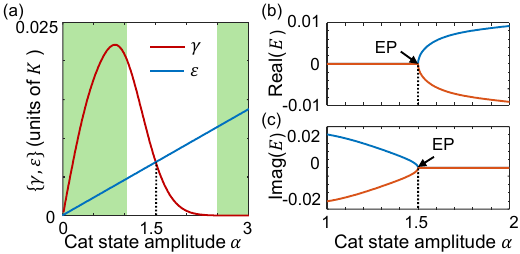}
		\caption{(a) The loss/gain rate $\gamma$ (red solid) and the coherent coupling $\epsilon$ (blue solid) as functions of the cat amplitude $\alpha$. Their intersection at $\alpha = 1.5$ defines the EP. The white region indicates where the discrepancy between the effective and full models is below $1\%$, validating our approach. (b) Real and (c) imaginary parts of the eigenvalues $E_{1,2}$ of $\mathcal{H}$ versus $\alpha$, clearly showing the $\mathcal{P}\mathcal{T}$-symmetry breaking transition at the EP. Parameters: $\kappa^{\phi}=0.05K$, $\Omega = 0.0023K$.}
		\label{fig3}
	\end{figure}

	\textit{$\mathcal{P}\mathcal{T}$-symmetry breaking in cat state subspace.}---The dynamics within the cat-state subspace ${\ket{\mathcal{C_{\pm}}}}$ is governed by the effective non-Hermitian Hamiltonian $\hat{\mathcal{H}}$ in Eq.~(\ref{eq14}), which takes the form of a $\mathcal{P}\mathcal{T}$-symmetric dimer. The off-diagonal element $\epsilon$ represents the coherent coupling between the even- and odd-parity cat states, while the imaginary part of the diagonal elements $\gamma$ quantify the parity-dependent loss and gain arising from the dephasing-induced \textit{asymmetric leakage}. Figure~\ref{fig3}(a) shows how $\gamma$ and $\epsilon$ vary with $\alpha$ when $\kappa^{\phi}, \Omega \ll K$, e.g., $\kappa^{\phi}=0.05K$ and $\Omega = 0.0023K$. 
	
	The Hamiltonian in Eq.~(\ref{eq14}) breaks Hermiticity and leads to eigenvalues those are generally complex, with their imaginary parts determining the amplification or decay of modal amplitudes during time evolution~\cite{Bender2007,ElGanainy2018,Ashida2020,Ashida2020,Bergholtz2021}. In such non-Hermitian systems, the conventional orthogonality of eigenvectors is replaced by biorthogonality~\cite{ElGanainy2018,Bergholtz2021}. The right eigenvectors $\ket{\phi_{j}}$ $(j=\pm)$, satisfying $\hat{\mathcal{H}}\ket{\phi_{j}} = E_{j}\ket{\phi_{j}}$, are complemented by left eigenvectors $\ket{\tilde{\phi}_{j}}$, defined through $\hat{\mathcal{H}}^{\dagger}\ket{\tilde{\phi}_{j}} = E_{j}^{*}\ket{\tilde{\phi}_{j}}$, which together satisfy the biorthogonality condition $\langle\tilde{\phi}_{i}|\phi_{j}\rangle = \delta_{ij}$ $(i=\pm)$. 
	
	A defining feature of non-Hermitian systems is the exceptional point (EP), which occurs at $\gamma = \epsilon$ (corresponding to $\alpha = 1.5$ in our setup). At the EP, the eigenvalues and the corresponding eigenvectors coalesce, leading to a non-diagonalizable Jordan block structure~\cite{ElGanainy2018,Miri2019}. This critical point separates two dynamically distinct phases: the exact $\mathcal{P}\mathcal{T}$-symmetric phase and the broken $\mathcal{P}\mathcal{T}$-symmetric phase~\cite{ElGanainy2018}. The EP is accessible simply by tuning the cat amplitude $\alpha$, a parameter that is straightforwardly controlled in Kerr-cat qubit experiments~\cite{Rglade2024}.
	
	The $\mathcal{P}\mathcal{T}$-symmetry breaking process is clearly displayed in Figs.~\ref{fig3}(b) and (c). For $\alpha>1.5$ ($\gamma<\epsilon$), the system resides in the $\mathcal{P}\mathcal{T}$-exact phase, characterized by entirely real eigenvalues $E_{\pm} = \pm \epsilon \cos\theta$, where $\theta = \sin^{-1}(\gamma/\epsilon)$. In this phase, the biorthogonal eigenvectors take the form $|\phi_{\pm}\rangle = [1, \pm e^{\mp i\theta}]^T$, supporting balanced, oscillatory dynamics without net amplification or decay—a signature of unbroken $\mathcal{P}\mathcal{T}$ symmetry~\cite{PhysRevLett.103.093902}.
	
	Conversely, when $\alpha<1.5$ ($\gamma>\epsilon$), the system enters the $\mathcal{P}\mathcal{T}$-broken phase, where the eigenvalues form complex conjugate pairs $E_{\pm} = \pm i\epsilon\sinh\theta$, with $\theta = \cosh^{-1}(\gamma/\epsilon)$. The corresponding eigenvectors $|\phi_{\pm}\rangle = [1, \pm ie^{\mp \theta}]^T$ become strongly non-orthogonal and exhibit spontaneous symmetry breaking: one eigenmode localizes predominantly on the ``gain'' site ($|\mathcal{C}_+\rangle$), experiencing exponential growth, while the other localizes on the ``loss'' site ($|\mathcal{C}_-\rangle$), undergoing an exponential decay~\cite{Rter2010}. This distinct localization, quantified by the markedly different magnitudes of the eigenvector components, is a hallmark of $\mathcal{P}\mathcal{T}$-symmetry breaking and has been observed in various photonic and superconducting platforms~\cite{Feng2012}. The transition between these phases, controlled by the cat amplitude $\alpha$, underscores the versatility of Kerr cat qubits for investigating fundamental non-Hermitian phenomena.

		\begin{figure}
		\centering
		\includegraphics[scale=1]{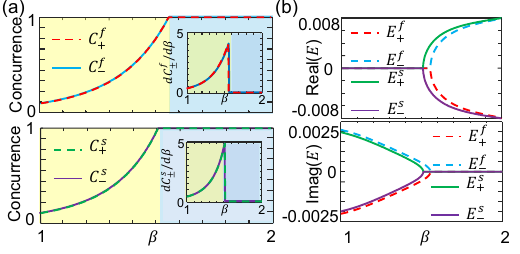}
		\caption{(a) Concurrence $C^{f,(s)}_{\pm}$ for the eigenstates $|\Phi^{f,(s)}_{\pm}\rangle$ as a function of $\beta$. Insets show the derivatives $dC^{f,(s)}_{\pm}/d\beta$ ($dC^{f,(s)}_{+}/d\beta$ are red and green dotted lines, respectively; $dC^{f,(s)}_{-}/d\beta$ are blue and purple solid lines, respectively), highlighting the critical behavior near the exceptional points. (b) Real and imaginary parts of the eigenvalues $E^{f,(s)}_{\pm}$ versus $\beta$, showing the coalescence of eigenvalues at the EPs. Parameters: $\alpha=2$, $\kappa^{\phi}_{1}=\kappa^{\phi}_{2}=0.05K_{1}$, $g=0.001K_{1}$, and $K_{2}=K_{1}$.}
		\label{fig4}
	    \end{figure}
	\textit{Entanglement phase transition in two cat qubits.}---Having established $\mathcal{P}\mathcal{T}$-symmetry breaking in a single cat qubit, we now demonstrate how coupling two such qubits enables the observation of a genuinely quantum non-Hermitian phenomenon: an entanglement phase transition induced by an exceptional point. Such transitions, where the entanglement changes non-analytically, have no direct counterpart in Hermitian quantum mechanics or classical physics~\cite{PhysRevLett.131.260201}. We consider two KPOs with intercavity coupling strength $g$. The interaction Hamiltonian is $\hat{H}_{I}=g(\hat{a}^{\dagger}\hat{b}+\hat{a}\hat{b}^{\dagger})$ under the rotating-wave approximation, where $\hat{a}$ and $\hat{b}$ ($\hat{a}^{\dagger}$ and $\hat{b}^{\dagger}$) are the annihilation (creation) operators of the first and second KPO, respectively. The Kerr nonlinearity, two-photon driving amplitude, cat-state amplitude, and dephasing rate for the $k$th cavity ($m=1,2$) are denoted as $K_m$, $P_m$, and $\kappa^{\phi}_m$, respectively. The cat size of the first and the second KPO are $\alpha$ and $\beta$, respectively. Projecting onto the joint cat-state subspace spanned by ${|\mathcal{C}_{\pm}\rangle_1 \otimes |\mathcal{C}_{\pm}\rangle_2}$, we obtain the effective non-Hermitian Hamiltonian (See in the Supplementary Material~\cite{YKX251014}):
	\begin{equation}
		\hat{H}_{\rm 2KPO}=
		\begin{pmatrix}
			i\Delta & 0 & 0 & J \\
			0 & i\delta & J & 0 \\
			0 & J & -i\delta & 0 \\
			J & 0 & 0 & -i\Delta
		\end{pmatrix},
		\label{eq15}
	\end{equation}
	where $\Delta = \gamma_1+\gamma_2$, $\delta = \gamma_1-\gamma_2$, $\gamma_k$ is the loss/gain rate of the $k$th cat qubit, and $J = 2\alpha\beta g$ represents the effective coupling between the qubits. This block-diagonal structure reflects the parity symmetry of the system, with the upper and lower blocks corresponding to different parity sectors. The eigenenergy and eigenstated of $\hat{H}_{\rm 2KPO}$ take the following form:

\begin{equation}
	\begin{aligned}
		E^{f}_{\pm} &= \pm \sqrt{J^{2} - \Delta^2}, \quad 
		E^{s}_{\pm} = \pm \sqrt{J^{2} - \delta^2}, \\
		\Phi^{f}_{\pm} &= \mathcal{N}^{f}_{\pm}[\mathcal{E}^{f}_{\pm},\ 0,\ 0,\ J]^{T}, \\
		\Phi^{s}_{\pm} &= \mathcal{N}^{s}_{\pm}[0,\ \mathcal{E}^{s}_{\pm},\ J,\ 0]^{T},
	\end{aligned}
\end{equation}
with $\mathcal{N}^{f,(s)}_{\pm} = (J^2 + \mathcal{E}^{f,(s)}_{\pm})^{-1/2}$, where $\mathcal{E}^{f}_{\pm} = i\Delta \pm \sqrt{J^{2} - \Delta^2}$ and $\mathcal{E}^{s}_{\pm} = i\delta \pm \sqrt{J^{2} - \delta^2}$. The superscripts $f$ and $s$ correspond to the first and second entangled subspaces, respectively.
	
	In contrast to conventional non-Hermitian systems, here each eigenenergy $E^{f,(s)}_{\pm}$ possessed by the two entangled components, neither of which has its own state. The nonclassicality of these eigenstates $\Phi^{f,(s)}_{\pm}$ is directly evidenced by the two body entanglement, which we quantify using concurrence~\cite{PhysRevLett.80.2245}
	\begin{equation}
		C^{f,(s)}_{\pm}=\frac{2J|\mathcal{E}^{f,(s)}_{\pm}|}{J^{2}+|\mathcal{E}^{f,(s)}_{\pm}|^2}.
	\end{equation}
	
	Figure~\ref{fig4}(a) demonstrates the entanglement phase transition controlled by the amplitude $\beta$ of the second cat qubit. For small $\beta$ ($\beta \approx 1$), all eigenstates exhibit near-zero entanglement ($C^{f,(s)}_{\pm} \approx 0$). As $\beta$ increases, the entanglement grows smoothly until reaching critical points where the derivatives $dC^{f,(s)}_{\pm}/d\beta$ (insets) diverge—a hallmark of non-analytic behavior characteristic of phase transitions. At their respective EPs, the eigenstate pairs $(|\Phi^{f}_{\pm}\rangle)$ and $(|\Phi^{s}_{\pm}\rangle)$ each coalesce into identical maximally entangled states with $C^{f,(s)}_{\pm} = 1$.
	
	The eigenvalue structure in Fig.~\ref{fig4}(b) confirms that these entanglement transitions coincide with EPs: $E^{f}_{+}$ and $E^{f}_{-}$ coalesce when $\Delta = J$, while $E^{s}_{+}$ and $E^{s}_{-}$ coalesce when $\delta = J$. This simultaneous coalescence of eigenvalues and eigenvectors, accompanied by the emergence of maximal entanglement, represents a purely quantum-mechanical manifestation of non-Hermitian criticality~\cite{PhysRevLett.131.260201}.

	\textit{Discussions and conclusions.}---Our proposal is experimentally grounded in established Kerr-cat qubit implementations. Recent works have realized stabilized cat states with amplitudes $\alpha$=1.5--2.5, and Kerr nonlinearities $K/2\pi \sim {\rm 6MHz}$ in superconducting circuits~\cite{Grimm2020,PhysRevX.14.041049}, while maintaining pure dephasing rates $\kappa^{\phi}\sim10^{-3}K$ that dominate the error budget~\cite{Lescanne2020,PhysRevX.15.011070}. Critically, the exponential suppression of bit-flip errors—demonstrated with bit-flip times exceeding 10 seconds~\cite{Rglade2024}—confirms that dephasing indeed emerges as the dominant noise channel in these systems, establishing the feasibility of observing our predicted non-Hermitian effects.
	
	In conclusion, we have shown that pure dephasing in a single Kerr-cat qubit induces non-Hermitian dynamics in its cat-state subspace ${\ket{\mathcal{C_{\pm}}}}$, enabling a $\mathcal{P}\mathcal{T}$-symmetry-breaking transition. The system parameter governing this transition, i.e., the cat amplitude $\alpha$, is highly and readily tunable in experiments. With two coupled cat qubits, this mechanism leads to an entanglement phase transition at exceptional points, which can be directly accessed simply by tuning the individual cat amplitudes $\alpha$ and $\beta$. Our results transform the unavoidable dephasing noise in cat qubits from a liability into a powerful resource for simulating non-Hermitian physics, thereby establishing coupled cat qubits as a versatile platform for exploring active non-Hermitian quantum phenomena and opening a new avenue for the study of open quantum systems.

	\textit{Acknowledgments.}---Y.-H.C. was supported by the National Natural Science Foundation of China (Grant Nos.~12304390 and 12574386), the Fujian 100 Talents Program, and the Fujian Minjiang Scholar Program. Z.-C.S. was supported by the National Natural Science Foundation of China (Grant No.~62571129) and the Natural Science Foundation of Fujian Province (Grant No.~2025J01456). Y.X. was supported by the National Natural Science Foundation of China (Grant Nos.~11575045 and 62471143), the Natural Science Funds for Distinguished Young Scholar of Fujian Province (Grant No.~2020J06011), and the Project from Fuzhou University (Grant No.~JG202001-2).

	\bibliography{reference}
\end{document}